\documentclass[aip, reprint, twocolumn, longbibliography, superscriptaddress]{revtex4-1}
\usepackage{fullpage}
\usepackage{graphicx}
\usepackage{dcolumn}
\usepackage{bm}
\usepackage{color}
\usepackage{amsmath, amsthm, amssymb, mathtools, cases}
\usepackage[modulo]{lineno}
\usepackage[usenames,dvipsnames]{xcolor}
\usepackage{ulem}
\usepackage{epstopdf}
\newcommand{\ba}{\begin{eqnarray}}
\newcommand{\ea}{\end{eqnarray}}

\usepackage{float}
\floatstyle{boxed}

\graphicspath{ {Figures/} }

\begin{document}
\title{Impact of contact resistance in Lorenz number measurements}
\author{Kin Chung Fong}
\email{fongkc@gmail.com}
\affiliation{Raytheon BBN Technologies, Quantum Information Processing Group, Cambridge, Massachusetts 02138, USA}
\date{\today}

\begin{abstract}
We analyze the effect of contact resistance on the Lorenz number measurement based on direct electronic thermal conductivity experiments. The contact resistance can significantly limit the experimental measured value when the Lorenz number is enhanced, but not as much so when it is suppressed, should the Wiedemann-Franz law be violated. The result provides the conditions of the potential false negative error and highlights the importance of improving the contact resistance in studying non-Fermi liquid behavior in thermal transport experiments.\end{abstract}
\maketitle

Considerable interest has arisen to understand the non-Fermi liquid behavior in strongly correlated systems by the Wiedemann-Franz law \cite{Crossno:2016iya, Gooth:2017ul, Lee:2017gg, Mahajan:2013fr, Wakeham:2011dp, Kim:2009eh, Smith:2008kv, Paglione:2006gt, Proust:2002hi, Hill:2001gf}. As the electrons are responsible for transporting both the charge and heat, the WF law states that the ratio of their conductivities should be proportional to the temperature $T$ with a proportional constant $\mathcal{L}$ called the Lorenz number:
$\kappa/\sigma=\mathcal{L}T$ where $\kappa$ and $\sigma$ are the thermal and electrical conductivity, respectively. The Lorenz number in Fermi liquid theory, $\mathcal{L}_0 = (\pi^2/3)(k_B/e)^2$, is a universal number containing only two fundamental constants, the Boltzmann constant $k_B$ and electrical charge $e$. Experimentally, this simple law holds remarkably well in wide ranges of carrier densities and masses \cite{Kumar:1993uk} as long as the collisions are elastic. This robustness makes the violation of the WF law a strong indicator for non-Fermi liquid behavior \cite{Mahajan:2013fr} if it is not trivially broken.

To test the WF law, the two major sources of errors are the additional thermal conductivity from phonons \cite{Lee:2017gg} and thermal boundary effects \cite{Wilson:2012dr, Majumdar:2004vx, Cahill:2014p011305}. To overcome these challenges, the new thermal conductivity techniques \cite{Volklein:2009kh, Fong:2012ut, Yigen:2013we} are advantagous in measuring the electronic thermal conductivity directly without the phonon contribution by probing the electron temperature rise with Joule heating applied to the sample in a two-terminal configuration. The electron temperature can be measured by either the resistive thermometry \cite{Volklein:2009kh, Yigen:2013we} or a high sensitivity setup called Johnson noise radiometer \cite{Fong:2012ut} for its similarity to the Dicke radiometer in measuring the temperature of the cosmic microwave background by coupling the blackbody radiation from space \cite{Dicke:1946vp}. The ratio of Joule heating to the electron temperature rise will provide the electronic thermal conductivity without the need of subtracting the phonon contribution that based usually on simulation. However, to the best of our knowledge, there is no quantitative analysis so far to understand how the inevitable electrical contact resistance would impact the Lorenz number measurement in these new methods. Specifically, would the experimental test give a false positive or negative error due to contact resistance? We will answer this question by calculating the Lorenz number value that is measured in an experiment as a function of the Lorenz number in the sample and contact resistance. Our analysis shows that the contact resistance in purely electronic thermal conductivity measurements can significantly limit the experimental value when the Lorenz number is enhanced, but not as much so when it is suppressed, should the WF law be violated.

\begin{figure}
\includegraphics[width=0.8\columnwidth]{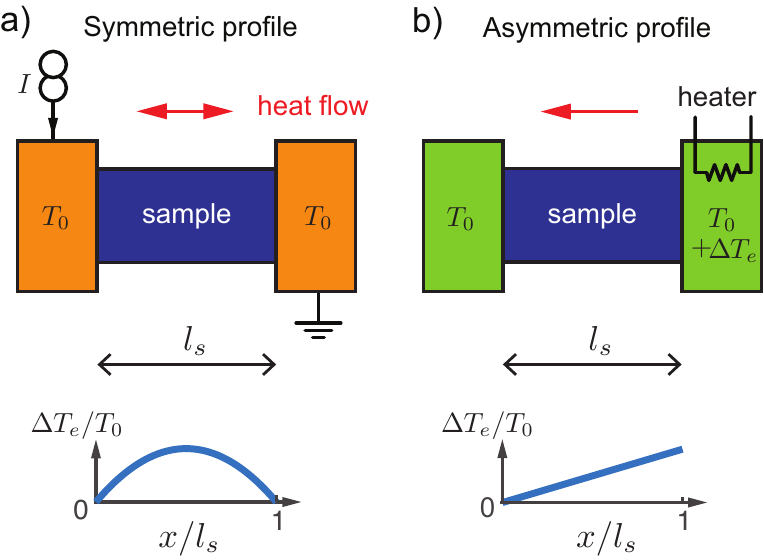}
\caption{Schematic diagrams of the two-terminal experimental setups to measure the electronic thermal conductivity in testing the Wiedemann-Franz law. In (a), the temperature rise is due to the current bias and its profile is symmetric. It can be measured using either the Johnson noise radiometer or resistive thermometry. In (b), the temperature rise is due to the heater outside the sample and its profile is asymmetric.}
\label{fig:Schematics}
\end{figure}

Fig. \ref{fig:Schematics}a shows the schematic diagram of the electronic thermal conductivity measurement setup to test the WF law in a sample connected with two metallic contacts on the sides \cite{Volklein:2009kh}. A thermal gradient is generated by Joule heating using a bias current $I$ flowing through the sample. The heat can be dissipated through the electronic diffusion and electron-phonon (EP) coupling, such that
\ba p_{J} = -\mathbf{\nabla}\cdot\left(\kappa\mathbf{\nabla}T_{\rm{e}}\right) - \Sigma(T_e^\delta - T_0^\delta)\text{,}\label{eqn:HeatTransferDiffEqn}\ea where $p_{J}$ is the Joule heating per unit volume, $\Sigma$ and $\delta$ are the EP coupling constant and temperature power law constant, respectively, and $T_e(x)$ is the elevated electron temperature at position $x$ of this sample from the base temperature $T_0$. We will only consider hereafter the experimental conditions in which the heat transfer by the electron diffusion is much larger than that by the EP coupling. This is valid at low enough temperatures because the EP coupling has a higher power law in temperature, i.e. $\delta \geq 3$ \cite{Wellstood:1994ec, Chen:2012et} than the WF law. Otherwise, the experimental analysis will need to include both the phonon thermal conductivity and phonon thermal boundary resistance, which is beyond the scope of this study.

In one-dimension with the EP heat channel neglected, the heat diffusion equation becomes: \ba P_{J} = -\frac{\mathcal{L}}{2R}l^2\frac{d^2}{dx^2}T_{\rm{e}}^2\label{eqn:OneDDiffEqn}\ea where $R$ and $l$ are the electrical resistance and distance between the two terminals, respectively, and $P_J = I^2R$ is the total Joule heating on the two-terminal device. The first boundary condition is $T_e(x=0) = T_e(x=l) = T_0$, as the electrons in the thick enough \cite{Wellstood:1994ec} metallic contacts are well-thermalized by both the electron diffusion and EP coupling; the second one is $\left.\partial T/\partial x\right\vert_{x=l/2} = 0$, as the heat flows out from the middle of the sample to the contacts in opposite directions. The heat diffusion equation results a quasi-parabolic $T_e$ given by:
\ba T_e^{\rm{(S)}}(x) &=& \sqrt{-\frac{I^2R^2}{\mathcal{L}}\left(\frac{x}{l}\right)^2+\frac{I^2R^2}{\mathcal{L}}\left(\frac{x}{l}\right)+T_0^2}\nonumber\\
&\simeq& T_0+\frac{1}{2}\frac{I^2R^2}{\mathcal{L}T_0}\left(\frac{x}{l}-\frac{x^2}{l^2}\right)\text{,}\label{eqn:TProfile2}\ea where the superscript (S) notates the symmetric temperature profile which is sketched in the lower panel in Fig. \ref{fig:Schematics}a. The approximation in Eqn. \ref{eqn:TProfile2} is made for a small temperature rise $\Delta T_e\equiv (T_{\rm{e}}(x)-T_0)\ll T_0$. This parabolic profile is the consequence of the linearly increasing heat flow from the center of the sample to the two metallic contacts (red arrow) because the bias current generates an uniform Joule heating along the sample.

The electronic thermal conductivity can also be measured in a similar setup with an uniform heat flow $P_h$ in one direction through the sample generated by a separated heater as shown in Fig. \ref{fig:Schematics}b. The temperature profile is linear, i.e. $T_e^{\rm{(A)}} = T_0+(P_hR/\mathcal{L}T_0)(x/l)$, where the superscript (A) notates the asymmetric temperature profile. It has the simplicity of no Joule heating generated in the sample due to current bias. However, the heat may leak out to locations in proximity to the heated metallic contact without passing through the sample. No purely electronic thermal conductivity has been measured by this method so far. We will focus more on the symmetric setup with the result of the asymmetric one for comparison.

In the experiments, the average electron temperature $\langle T_e\rangle$ is measured. Specifically, in the Johnson noise radiometer, $\langle T_e\rangle = \int \rho(x) T_e(x) dx/ \int \rho(x) dx$ where $\rho(x)$ is the position dependent resistivity. This is because the Johnson noise power fluctuation is proportional to $4Rk_BT_0$. In the resistive thermometry, $\langle T_e\rangle$ has the same form if the temperature coefficient is a constant along the entire device. For a homogeneous sample, the integration reduces to spatial average and the results are: 
\begin{subequations}\begin{align} \langle T_e^{\rm{(S)}}\rangle &= T_0+\frac{1}{12}\frac{P_{J}R}{\mathcal{L}T_0}\\
\langle T_e^{\rm{(A)}}\rangle &= T_0+\frac{1}{2}\frac{P_hR}{\mathcal{L}T_0}.\end{align}\end{subequations} The numerical prefactor in the second terms are due to the integration of the temperature profiles. Interestingly, this factor is universal for an arbitrary shaped sample in the two-terminal setup \cite{Brian}. Using Eqn. (4), the Lorenz number obtained in the experiments are $\mathcal{L}_m^{\rm{(S)}} \equiv P_JR/(12T_0\langle\Delta T_e^{\rm{(S)}} \rangle)$ for the symmetric profile, and $\mathcal{L}_m^{\rm{(A)}} \equiv P_hR/(2T_0\langle\Delta T_e^{\rm{(A)}} \rangle)$ for the asymmetric profile.

\begin{figure}
\includegraphics[width=0.8\columnwidth]{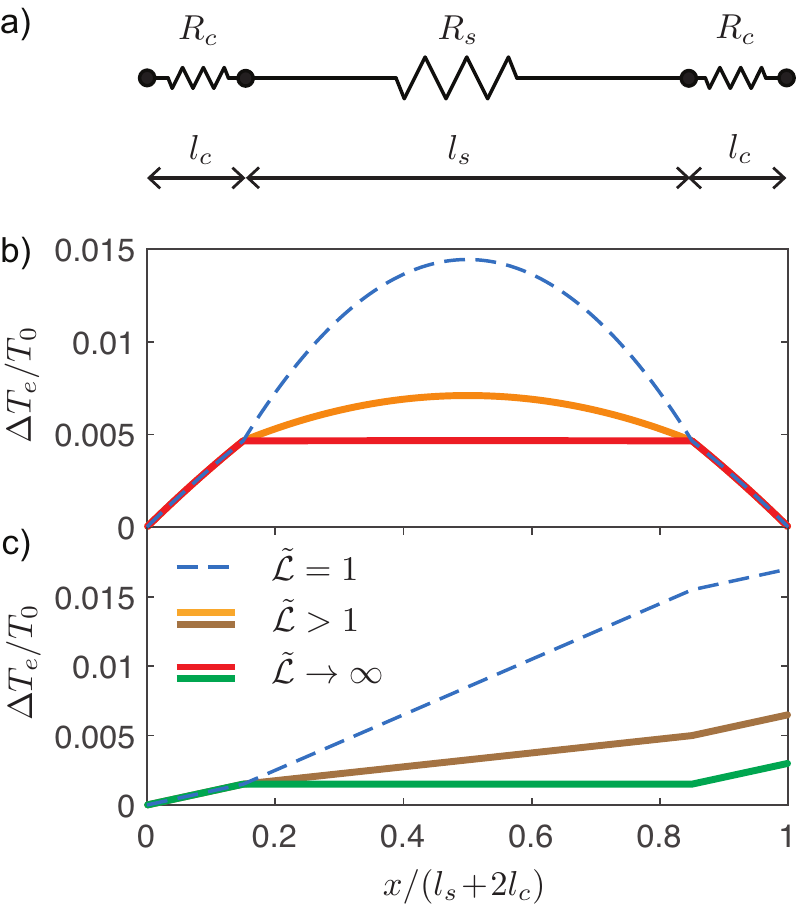}
\caption{a) The sample resistance in series with the contact resistance in the two-terminal setup. Their thermal resistances are also connected in series. (b) Symmetric (Eqn.~6) and (c) asymmetric temperature profiles with a finite contact resistance for $\mathcal{L}=\mathcal{L}_0$ (dashed lines) in the sample using $R_c/R_s = 0.5$ and $l_c/l_s = 0.15/0.7$. The temperature gradient across the sample is reduced when $\mathcal{L}>\mathcal{L}_0$ (solid lines).}
\label{fig:TProfile}
\end{figure}

We can analyze the error in the Lorenz number measurement by modeling the temperature profile with a finite contact resistance connected in series with the sample. Denoting the contact resistance of each contact as $R_c$, the total resistance would be $R = 2R_c + R_s$ where $R_s$ is the resistance of the sample alone (see Fig. \ref{fig:TProfile}a). Both theoretical \cite{Mahan:1999eg} and experimental \cite{Gundrum:2005cx, Wilson:2012dr} studies have shown that the electronic thermal conductivity at the sharp contact interface can be understood as a diffusive process and the WF law is valid at the contact. To model the temperature rise at the contact, we assume the contact resistance takes place in a short, hypothetical distance $l_c$ such that $l = l_s+2l_c$ with $l_s$ being the sample length along the direction of current flow. We emphasis that $l_c$ is a modeling construction and our final result depends only on the resistance ratio of the contact and sample, not on this hypothetical distance. We shall show later that this assumption is equivalent to model, alternatively, the contact resistance with a discrete temperature drop across the interface with the Joule heating of the contact resistance dissipates via the two ends of the contact. We note that Peltier effect \cite{Grosse:2011ka, VeraMarun:2016ih} may occur at the contact. However, as it scales as $T_0^2$, the heat flow is dominated by the thermal conductivity at low temperatures. The resistivity of the contact $\rho_c$ and sample $\rho_s$ are $\rho_c=R_cA/l_c$ and $\rho_s=R_sA/l_s$, where $A$ is the cross-sectional area, respectively.

We can solve the heat diffusion equation, Eqn. (\ref{eqn:OneDDiffEqn}), for the three sessions device composed of the sample sandwiched between two contact resistors in which $\mathcal{L}=\mathcal{L}_0$. Firstly, from the similarity of the boundary conditions in obtaining Eqn. (\ref{eqn:TProfile2}), we can write $T_e^{\rm{(S)}}(x)$ within the sample, i.e. $l_c<x<l_s/2+l_c$, as: \begin{equation} T_e^{\rm{(S)}}(x) = \sqrt{-\frac{I^2R_s^2}{\mathcal{L}}\left(\frac{x-l_c}{l_s}\right)^2+\frac{I^2R_s^2}{\mathcal{L}}\left(\frac{x-l_c}{l_s}\right)+T_{cs}^2}\end{equation}, where $T_{cs} = T_e^{\rm{(S)}}(x=l_c) $ is the temperature at the contact and sample interface. Then, $T_e^{\rm{(S)}}(x) $ in the contact can be determined by the boundary conditions of the temperature and the contact-sample interfacial heat flow:
\begin{subequations} \label{eqn:BoundaryConditions}\begin{align} 
T_e^{\rm{(S)}}(x=0) &= T_0\\
\frac{I^2R_s}{2} &= \frac{A\mathcal{L}_0T_0}{\rho_c}\left .\frac{\partial T_e^{\rm{(S)}}}{\partial x}\right\vert_{x=l_c}
\end{align}\end{subequations} alongside the heat diffusion equation in the contact, i.e. $I^2\rho_c/A^2 = (\partial^2(T_c^{\rm{(S)}})^2/\partial x^2)\mathcal{L}_0/(2\rho_c)$.

Using $\Delta T_e \ll T_0$ (calculation detailed in Appendix), the solution of the symmetric temperature profile is:
\begin{widetext}\begin{equation}\label{eqn:TProfile3}
T_e^{\rm{(S)}}(x) \simeq 
  \begin{dcases*}
	\sqrt{-\frac{I^2R_c^2}{\mathcal{L}_0}\left(\frac{x}{l_c}\right)^2+\frac{I^2R_cR}{\mathcal{L}_0}\left(\frac{x}{l_c}\right)+T_0^2} & for $0\leq x \leq l_c$ (in the contact), \\
	T_e^{\rm{(S)}}(l-x) & for $l_s+l_c \leq x \leq l_s+2l_c$ (in the contact), \\
	\sqrt{-\frac{I^2R_s^2}{\mathcal{L}}\left(\frac{x-l_c}{l_s}\right)^2+\frac{I^2R_s^2}{\mathcal{L}}\left(\frac{x-l_c}{l_s}\right)+T_0^2\left(1+\frac{I^2R_c}{\mathcal{L}_0T_0^2}\frac{R_s+R_c}{2}\right)^2}& for $l_c\leq x\leq l_s+l_c$ (in the sample).
  \end{dcases*}\end{equation}\end{widetext}
The dashed line in Fig.  \ref{fig:TProfile}b plot the temperature profile for $\mathcal{L}=\mathcal{L}_0$ in the sample, with an arbitrarily chosen $R_c$ and $\rho_c$ ($l_c$) for illustration. If the WF law is violated in the sample, the temperature rise is lower than that of the dashed line for $\mathcal{L}>\mathcal{L}_0$, vice versa. For the limiting case when $\mathcal{L}\rightarrow\infty$, there is no temperature rise in the sample (red line).
 
Alternatively, we can model the contact-sample interface with a discrete temperature drop. Using the WF law, the thermal conductivity of the contact is $\mathcal{L}_0T_0/R_c$. The Joule heating $I^2R_c$ generated at the contact due to the current bias can be distributed equally on either side of the contact-sample interface so that the heat flow through the contact is $I^2(R_s+R_c)/2$. Therefore, the discrete interfacial temperature drop is $I^2R_c(R_s+R_c)/ 2\mathcal{L}_0T_0$. This is same as $T_{cs}$ in the diffusive contact model using Eqn. (\ref{eqn:TProfile3}). For a small $R_c/R_s$ ratio, both models predict the same Lorenz number.

Similarly, the temperature profiles with a non-zero contact resistance in the asymmetric setup are plotted in Fig. \ref{fig:TProfile}c. The thermal conductance across the sample and contact resistance in the asymmetric profile are, to the first order, $\mathcal{L}T_0/R_s$ and $\mathcal{L}_0T_0/R_c$, respectively. The corresponding averaged temperature rises $\langle\Delta T_e^{\rm{(A)}}\rangle = \Delta T_e^{\rm{(A)}}(x=l_s/2+l_c) = P_hR_s/2\mathcal{L}T_0 + P_hR_c/\mathcal{L}_0T_0$.

To compare the measured Lorenz number in the experiment to the actual value in the sample, we can integrate the temperature profiles to get the measured Lorenz numbers, i.e.
\begin{subequations}\begin{align}\tilde{\mathcal{L}}_{\rm{m}}^{\rm{(S)}} &= \left[1-\left(1-\frac{1}{\tilde{\mathcal{L}}}\right)\left(\frac{1}{1+2\tilde{r}}\right)^3\right]^{-1} \label{eq:LmeasRadiometer}\\
\tilde{\mathcal{L}}_{\rm{m}}^{\rm{(A)}} &= \left(\frac{1+2\tilde{r}}{1+2\tilde{r}\tilde{\mathcal{L}}}\right)\tilde{\mathcal{L}}, \label{eq:LmeasSB}\end{align}\end{subequations} where $\tilde{r} = R_c/R_s$, $\tilde{\mathcal{L}} = \mathcal{L}/\mathcal{L}_0$ and $\tilde{\mathcal{L}}_{m}^{\rm{(k)}} = \mathcal{L}_m^{\rm{(k)}}/\mathcal{L}_0$ with (k) = (S) or (A). These two equations are the central results of this analysis. They express the Lorenz number that we would measure in experiments in terms of the contact resistance and the sample Lorenz number we intend to find.

\begin{figure}
\includegraphics[width=0.8\columnwidth]{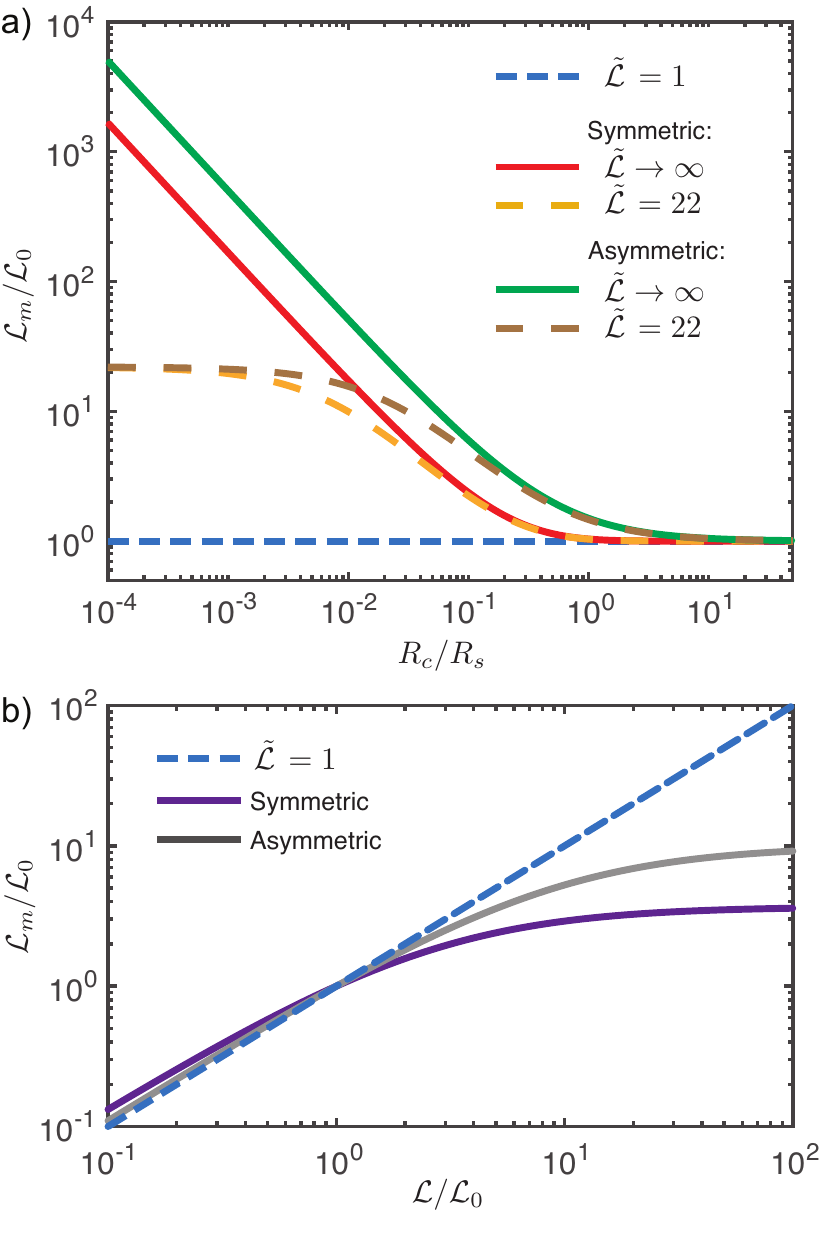}
\caption{a) The experimentally measured Lorenz ratio versus the resistance ratio of the contact and sample as described by Eqn. (8). (b) The experimentally measured Lorenz ratioversus the actual Lorenz ratio in the sample assuming the contact resistance is 10\% of the two-terminal resistance.}
\label{fig:MaxLorenz}
\end{figure}

We plot $\tilde{\mathcal{L}}_{m}$ for some experimental scenarios in Fig. \ref{fig:MaxLorenz}a. The blue dashed line shows the trivial case for $\tilde{\mathcal{L}}=1$ so that $\tilde{\mathcal{L}_{m}}=1$ for any contact resistance. For the limiting case when the sample $\tilde{\mathcal{L}}\rightarrow\infty$ (red), $\tilde{\mathcal{L}}_{m}^{(S)}$ grows from 1 towards the true sample $\tilde{\mathcal{L}}$ as the contact resistance decreases. This represents the maximum Lorenz number that the setup can measure as function of the contact resistance. We also plot $\tilde{\mathcal{L}}_{m}^{(S)}$ for $\tilde{\mathcal{L}} = 22$ (orange). The measured Lorenz number approaches to the true value only for small enough $R_c/R_s$ ratio, i.e. $\tilde{r} \ll 1/6\tilde{\mathcal{L}}$.

We can justify our model by comparing to the experimental data in the breakdown of WF law in graphene \cite{Crossno:2016iya} in which the largest reported $\mathcal{L}_{m}$ is 22$\mathcal{L}_0$. Using Eqn. (\ref{eq:LmeasRadiometer}), the upper bound of the contact resistance in the experiment is estimated to be $\tilde{r} \simeq 0.008$. With the two-terminal resistance near the charge neutrality of about 1100 $\Omega$, the contact resistance is about 9~$\Omega$ on each of the 9~$\mu$m wide contact. This corresponds to a contact resistance of about 81~$\Omega\cdot\mu$m. It is smaller but consistent to the measured value based on the transfer-length method on samples that are fabricated using the same method \cite{Wang:2013ch}. This points towards the importance of reducing the contact resistance \cite{Leonard:2011dh} for accurately determining $\mathcal{L}$ in the electron thermal conductivity measurement.

Fig. \ref{fig:MaxLorenz}a also plots the $\tilde{\mathcal{L}}_{m}^{(A)}$ for the asymmetric temperature profile excitation. The general dependence on the contact resistance in $\tilde{\mathcal{L}}_{m}^{(A)}$ is similar to that in $\tilde{\mathcal{L}}_{m}^{(S)}$, but the maximum measurable Lorenz number in the asymmetric profile is 3 times larger than that in its symmetric counterpart for a given $R_c/R_s$ ratio. Since the Lorenz number is deduced from the temperature rise for a given bias current, the higher elevated temperature in the linear profile is more advantageous than that in the parabolic one. For instance, the middle of the sample has no temperature gradient to yield any information in the symmetric setup.

To understand the impact of contact resistance when $\tilde{\mathcal{L}} < 1$, Fig. \ref{fig:MaxLorenz}b plots $\mathcal{L}_{m}$ as function of the Lorenz number in the sample using $\tilde{r} = 0.056$, i.e. $R_c = 0.1R$. For both temperature profiles, the measured Lorenz number does not diverge from the true sample value when $\tilde{\mathcal{L}} < 1$. This is because a stronger Lorenz number suppression corresponding to a smaller thermal conductivity in the sample. The temperature rises relatively higher in the sample than the contact, in effect diminishing the error due to the existence of contact resistance. 

In this report, we analyze the Lorenz number measurement error due to the contact resistance in the direct electronic thermal conductivity experimental setup. We find the measured Lorenz number will be upper bounded at the value of $R_s\mathcal{L}_0/6R_c$ by the existence of a contact resistance. The result shows that the contact resistance can drastically limit the measured value when the Lorenz number is enhanced, but not as much so when it is suppressed. This is because the contact resistance is connected in series with the sample along the direction of the heat flow and the temperature gradient is larger across the larger thermal resistor. When the Lorenz number grows infinitely, the temperature only rises across the contact resistance, imposing an experimental measurement limit which we calculate in this report. This result performs the due-diligence and highlights the importance of improving the contact resistance in WF law experiments. Otherwise the test of the WF law may obtain a false negative result, especially when the temperature sensitivity is limited. For graphene-metal contacts, the estimated contact resistance in the relativistic hydrodynamic experiment is still much larger than the theoretical limit \cite{Matsuda:2010fv}. This suggests there can be a lot of room for improvements in the future to understand the non-Fermi liquid behavior using thermal transport experiments.

The author thanks discussions with J. Sanchez-Yamagishi, B. Skinner, L. Levitov, J. Ravichandran, J. Waissman, J. Crossno, and P. Kim. This research was funded by Army Research Office under Cooperative Agreement Number W911NF-17-2-0086 and Raytheon BBN Technologies. K.C.F. acknowledges support from the Simons Center for Geometry and Physics, Stony Brook University at which some of the research for this paper was performed.

\appendix*
\section{Solving the heat diffusion equation in the contact}
Similar to the solution given by Eqn. (\ref{eqn:TProfile2}), the general solution to the heat diffusion equation, Eqn. (\ref{eqn:OneDDiffEqn}), in the contact is \ba T_e^{\rm{(S)}}(0\leq x\leq l_c) = \sqrt{D_cx^2+E_cx+F_0}.\ea Using the boundary conditions in Eqn. (\ref{eqn:BoundaryConditions}), we can obtain:
$D_c = -I^2R_c^2/(\mathcal{L}_0l_c^2)$, $F_0=T_0^2$, and \begin{widetext}\begin{equation}\label{eqn:CoeffEc}
E_c = \frac{I^2R_c}{2\mathcal{L}_0^2T_0^2}\left[R_c(4\mathcal{L}_0T_0^2+I^2R_s^2)+R_s\sqrt{4\mathcal{L}_0^2T_0^4+4\mathcal{L}_0T_0^2\cdot I^2R_c^2+(I^2R_sR_c)^2}\right]\left(\frac{1}{l_c}\right).\end{equation}\end{widetext}
We can simplify this exact solution of $T_e^{\rm{(S)}}(x)$ using \ba\frac{I^2R_c^2}{\mathcal{L}_0T_0^2}\text{,~}\frac{I^2R_c^2}{\mathcal{L}_0T_0^2}\ll 1\ea because the raised temperature $\Delta T_e$ is much smaller than $T_0$. Therefore, \ba E_c \simeq \frac{I^2R_cR}{\mathcal{L}_0}\left(\frac{1}{l_c}\right) \ea to the first order of $\Delta T_e/T_0$.

\bibliographystyle{apsrev}

\begin{thebibliography}{28}
\expandafter\ifx\csname natexlab\endcsname\relax\def\natexlab#1{#1}\fi
\expandafter\ifx\csname bibnamefont\endcsname\relax
  \def\bibnamefont#1{#1}\fi
\expandafter\ifx\csname bibfnamefont\endcsname\relax
  \def\bibfnamefont#1{#1}\fi
\expandafter\ifx\csname citenamefont\endcsname\relax
  \def\citenamefont#1{#1}\fi
\expandafter\ifx\csname url\endcsname\relax
  \def\url#1{\texttt{#1}}\fi
\expandafter\ifx\csname urlprefix\endcsname\relax\def\urlprefix{URL }\fi
\providecommand{\bibinfo}[2]{#2}
\providecommand{\eprint}[2][]{\url{#2}}

\bibitem[{\citenamefont{Crossno et~al.}(2016)\citenamefont{Crossno, Shi, Wang,
  Liu, Harzheim, Lucas, Sachdev, Kim, Taniguchi, Watanabe
  et~al.}}]{Crossno:2016iya}
\bibinfo{author}{\bibfnamefont{J.}~\bibnamefont{Crossno}},
  \bibinfo{author}{\bibfnamefont{J.~K.} \bibnamefont{Shi}},
  \bibinfo{author}{\bibfnamefont{K.}~\bibnamefont{Wang}},
  \bibinfo{author}{\bibfnamefont{X.}~\bibnamefont{Liu}},
  \bibinfo{author}{\bibfnamefont{A.}~\bibnamefont{Harzheim}},
  \bibinfo{author}{\bibfnamefont{A.}~\bibnamefont{Lucas}},
  \bibinfo{author}{\bibfnamefont{S.}~\bibnamefont{Sachdev}},
  \bibinfo{author}{\bibfnamefont{P.}~\bibnamefont{Kim}},
  \bibinfo{author}{\bibfnamefont{T.}~\bibnamefont{Taniguchi}},
  \bibinfo{author}{\bibfnamefont{K.}~\bibnamefont{Watanabe}},
  \bibinfo{author}{\bibfnamefont{T.~A.}~\bibnamefont{Ohki}},
  \bibinfo{author}{\bibfnamefont{K.~C.}~\bibnamefont{Fong}},
\bibinfo{journal}{Science}
  \textbf{\bibinfo{volume}{351}}, \bibinfo{pages}{1058} (\bibinfo{year}{2016}).

\bibitem[{\citenamefont{Gooth et~al.}(2017)\citenamefont{Gooth, Menges,
  Shekhar, S{\"u}{\ss}, Kumar, Sun, Drechsler, Zierold, Felser, and
  Gotsmann}}]{Gooth:2017ul}
\bibinfo{author}{\bibfnamefont{J.}~\bibnamefont{Gooth}},
  \bibinfo{author}{\bibfnamefont{F.}~\bibnamefont{Menges}},
  \bibinfo{author}{\bibfnamefont{C.}~\bibnamefont{Shekhar}},
  \bibinfo{author}{\bibfnamefont{V.}~\bibnamefont{S{\"u}{\ss}}},
  \bibinfo{author}{\bibfnamefont{N.}~\bibnamefont{Kumar}},
  \bibinfo{author}{\bibfnamefont{Y.}~\bibnamefont{Sun}},
  \bibinfo{author}{\bibfnamefont{U.}~\bibnamefont{Drechsler}},
  \bibinfo{author}{\bibfnamefont{R.}~\bibnamefont{Zierold}},
  \bibinfo{author}{\bibfnamefont{C.}~\bibnamefont{Felser}}, \bibnamefont{and}
  \bibinfo{author}{\bibfnamefont{B.}~\bibnamefont{Gotsmann}},
\eprint{arXiv:1706.05925} (\bibinfo{year}{2017}).

\bibitem[{\citenamefont{Lee et~al.}(2017)\citenamefont{Lee, Hippalgaonkar,
  Yang, Hong, Ko, Suh, Liu, Wang, Urban, Zhang et~al.}}]{Lee:2017gg}
\bibinfo{author}{\bibfnamefont{S.}~\bibnamefont{Lee}},
  \bibinfo{author}{\bibfnamefont{K.}~\bibnamefont{Hippalgaonkar}},
  \bibinfo{author}{\bibfnamefont{F.}~\bibnamefont{Yang}},
  \bibinfo{author}{\bibfnamefont{J.}~\bibnamefont{Hong}},
  \bibinfo{author}{\bibfnamefont{C.}~\bibnamefont{Ko}},
  \bibinfo{author}{\bibfnamefont{J.}~\bibnamefont{Suh}},
  \bibinfo{author}{\bibfnamefont{K.}~\bibnamefont{Liu}},
  \bibinfo{author}{\bibfnamefont{K.}~\bibnamefont{Wang}},
  \bibinfo{author}{\bibfnamefont{J.~J.} \bibnamefont{Urban}},
  \bibinfo{author}{\bibfnamefont{X.}~\bibnamefont{Zhang}},
    \bibinfo{author}{\bibfnamefont{C.}~\bibnamefont{Dames}},
      \bibinfo{author}{\bibfnamefont{S.~A.}~\bibnamefont{Hartnoll}},
        \bibinfo{author}{\bibfnamefont{O.}~\bibnamefont{Delaire}},
          \bibinfo{author}{\bibfnamefont{J.}~\bibnamefont{Wu}},
  \bibinfo{journal}{Science}
  \textbf{\bibinfo{volume}{355}}, \bibinfo{pages}{371} (\bibinfo{year}{2017}).

\bibitem[{\citenamefont{Mahajan et~al.}(2013)\citenamefont{Mahajan, Barkeshli,
  and Hartnoll}}]{Mahajan:2013fr}
\bibinfo{author}{\bibfnamefont{R.}~\bibnamefont{Mahajan}},
  \bibinfo{author}{\bibfnamefont{M.}~\bibnamefont{Barkeshli}},
  \bibnamefont{and} \bibinfo{author}{\bibfnamefont{S.~A.}
  \bibnamefont{Hartnoll}}, \bibinfo{journal}{Phys. Rev. B}
  \textbf{\bibinfo{volume}{88}}, \bibinfo{pages}{125107}
  (\bibinfo{year}{2013}).

\bibitem[{\citenamefont{Wakeham et~al.}(2011)\citenamefont{Wakeham, Bangura,
  Xu, and Mercure}}]{Wakeham:2011dp}
\bibinfo{author}{\bibfnamefont{N.}~\bibnamefont{Wakeham}},
  \bibinfo{author}{\bibfnamefont{A.~F.} \bibnamefont{Bangura}},
  \bibinfo{author}{\bibfnamefont{X.}~\bibnamefont{Xu}}, \bibnamefont{and}
  \bibinfo{author}{\bibfnamefont{J.~F.} \bibnamefont{Mercure}},
  \bibinfo{journal}{Nat. Commun.} \textbf{\bibinfo{volume}{2}},
  \bibinfo{pages}{396} (\bibinfo{year}{2011}).

\bibitem[{\citenamefont{Kim and P{\'e}pin}(2009)}]{Kim:2009eh}
\bibinfo{author}{\bibfnamefont{K.~S.} \bibnamefont{Kim}} \bibnamefont{and}
  \bibinfo{author}{\bibfnamefont{C.}~\bibnamefont{P{\'e}pin}},
  \bibinfo{journal}{Phys. Rev. Lett.} \textbf{\bibinfo{volume}{102}},
  \bibinfo{pages}{156404} (\bibinfo{year}{2009}).

\bibitem[{\citenamefont{Smith et~al.}(2008)\citenamefont{Smith, Sutherland,
  Lonzarich, Saxena, Kimura, Takashima, Nohara, and Takagi}}]{Smith:2008kv}
\bibinfo{author}{\bibfnamefont{R.~P.} \bibnamefont{Smith}},
  \bibinfo{author}{\bibfnamefont{M.}~\bibnamefont{Sutherland}},
  \bibinfo{author}{\bibfnamefont{G.~G.} \bibnamefont{Lonzarich}},
  \bibinfo{author}{\bibfnamefont{S.~S.} \bibnamefont{Saxena}},
  \bibinfo{author}{\bibfnamefont{N.}~\bibnamefont{Kimura}},
  \bibinfo{author}{\bibfnamefont{S.}~\bibnamefont{Takashima}},
  \bibinfo{author}{\bibfnamefont{M.}~\bibnamefont{Nohara}}, \bibnamefont{and}
  \bibinfo{author}{\bibfnamefont{H.}~\bibnamefont{Takagi}},
  \bibinfo{journal}{Nature} \textbf{\bibinfo{volume}{455}},
  \bibinfo{pages}{1220} (\bibinfo{year}{2008}).

\bibitem[{\citenamefont{Paglione et~al.}(2006)\citenamefont{Paglione, Tanatar,
  Hawthorn, Ronning, Hill, Sutherland, Taillefer, and
  Petrovic}}]{Paglione:2006gt}
\bibinfo{author}{\bibfnamefont{J.}~\bibnamefont{Paglione}},
  \bibinfo{author}{\bibfnamefont{M.~A.} \bibnamefont{Tanatar}},
  \bibinfo{author}{\bibfnamefont{D.~G.} \bibnamefont{Hawthorn}},
  \bibinfo{author}{\bibfnamefont{F.}~\bibnamefont{Ronning}},
  \bibinfo{author}{\bibfnamefont{R.~W.} \bibnamefont{Hill}},
  \bibinfo{author}{\bibfnamefont{M.}~\bibnamefont{Sutherland}},
  \bibinfo{author}{\bibfnamefont{L.}~\bibnamefont{Taillefer}},
  \bibnamefont{and} \bibinfo{author}{\bibfnamefont{C.}~\bibnamefont{Petrovic}},
  \bibinfo{journal}{Phys. Rev. Lett.} \textbf{\bibinfo{volume}{97}},
  \bibinfo{pages}{106606} (\bibinfo{year}{2006}).

\bibitem[{\citenamefont{Proust et~al.}(2002)\citenamefont{Proust, Boaknin,
  Hill, Taillefer, and Mackenzie}}]{Proust:2002hi}
\bibinfo{author}{\bibfnamefont{C.}~\bibnamefont{Proust}},
  \bibinfo{author}{\bibfnamefont{E.}~\bibnamefont{Boaknin}},
  \bibinfo{author}{\bibfnamefont{R.~W.} \bibnamefont{Hill}},
  \bibinfo{author}{\bibfnamefont{L.}~\bibnamefont{Taillefer}},
  \bibnamefont{and} \bibinfo{author}{\bibfnamefont{A.~P.}
  \bibnamefont{Mackenzie}}, \bibinfo{journal}{Phys. Rev. Lett.}
  \textbf{\bibinfo{volume}{89}}, \bibinfo{pages}{147003}
  (\bibinfo{year}{2002}).

\bibitem[{\citenamefont{Hill et~al.}(2001)\citenamefont{Hill, Proust,
  Taillefer, Fournier, and Greene}}]{Hill:2001gf}
\bibinfo{author}{\bibfnamefont{R.~W.} \bibnamefont{Hill}},
  \bibinfo{author}{\bibfnamefont{C.}~\bibnamefont{Proust}},
  \bibinfo{author}{\bibfnamefont{L.}~\bibnamefont{Taillefer}},
  \bibinfo{author}{\bibfnamefont{P.}~\bibnamefont{Fournier}}, \bibnamefont{and}
  \bibinfo{author}{\bibfnamefont{R.~L.} \bibnamefont{Greene}},
  \bibinfo{journal}{Nature} \textbf{\bibinfo{volume}{414}},
  \bibinfo{pages}{711} (\bibinfo{year}{2001}).

\bibitem[{\citenamefont{Kumar et~al.}(1993)\citenamefont{Kumar, Prasad, and
  Pohl}}]{Kumar:1993uk}
\bibinfo{author}{\bibfnamefont{G.~S.} \bibnamefont{Kumar}},
  \bibinfo{author}{\bibfnamefont{G.}~\bibnamefont{Prasad}}, \bibnamefont{and}
  \bibinfo{author}{\bibfnamefont{R.~O.} \bibnamefont{Pohl}},
  \bibinfo{journal}{J. Mater. Sci.} \textbf{\bibinfo{volume}{28}},
  \bibinfo{pages}{4261} (\bibinfo{year}{1993}).

\bibitem[{\citenamefont{Wilson and Cahill}(2012)}]{Wilson:2012dr}
\bibinfo{author}{\bibfnamefont{R.~B.} \bibnamefont{Wilson}} \bibnamefont{and}
  \bibinfo{author}{\bibfnamefont{D.~G.} \bibnamefont{Cahill}},
  \bibinfo{journal}{Phys. Rev. Lett.} \textbf{\bibinfo{volume}{108}},
  \bibinfo{pages}{255901} (\bibinfo{year}{2012}).

\bibitem[{\citenamefont{Majumdar and Reddy}(2004)}]{Majumdar:2004vx}
\bibinfo{author}{\bibfnamefont{A.}~\bibnamefont{Majumdar}} \bibnamefont{and}
  \bibinfo{author}{\bibfnamefont{P.}~\bibnamefont{Reddy}},
  \bibinfo{journal}{Appl. Phys. Letts.} \textbf{\bibinfo{volume}{84}},
  \bibinfo{pages}{4768} (\bibinfo{year}{2004}).

\bibitem[{\citenamefont{Cahill et~al.}(2014)\citenamefont{Cahill, Braun, Chen,
  Clarke, Fan, Goodson, Keblinski, King, Mahan, Majumdar
  et~al.}}]{Cahill:2014p011305}
\bibinfo{author}{\bibfnamefont{D.}~\bibnamefont{Cahill}},
  \bibinfo{author}{\bibfnamefont{P.}~\bibnamefont{Braun}},
  \bibinfo{author}{\bibfnamefont{G.}~\bibnamefont{Chen}},
  \bibinfo{author}{\bibfnamefont{D.}~\bibnamefont{Clarke}},
  \bibinfo{author}{\bibfnamefont{S.}~\bibnamefont{Fan}},
  \bibinfo{author}{\bibfnamefont{K.}~\bibnamefont{Goodson}},
  \bibinfo{author}{\bibfnamefont{P.}~\bibnamefont{Keblinski}},
  \bibinfo{author}{\bibfnamefont{W.}~\bibnamefont{King}},
  \bibinfo{author}{\bibfnamefont{G.}~\bibnamefont{Mahan}},
  \bibinfo{author}{\bibfnamefont{A.}~\bibnamefont{Majumdar}},
  \bibinfo{author}{\bibfnamefont{H.~J.}~\bibnamefont{Maris}},
  \bibinfo{author}{\bibfnamefont{S.~R.}~\bibnamefont{Phillpot}},
  \bibinfo{author}{\bibfnamefont{E.}~\bibnamefont{Pop}},
  \bibinfo{author}{\bibfnamefont{L.}~\bibnamefont{Shi}},
  \bibinfo{journal}{Appl. Phys. Rev.}
  \textbf{\bibinfo{volume}{1}}, \bibinfo{pages}{011305} (\bibinfo{year}{2014}).

\bibitem[{\citenamefont{V{\"o}lklein et~al.}(2009)\citenamefont{V{\"o}lklein,
  Reith, Cornelius, Rauber, and Neumann}}]{Volklein:2009kh}
\bibinfo{author}{\bibfnamefont{F.}~\bibnamefont{V{\"o}lklein}},
  \bibinfo{author}{\bibfnamefont{H.}~\bibnamefont{Reith}},
  \bibinfo{author}{\bibfnamefont{T.~W.} \bibnamefont{Cornelius}},
  \bibinfo{author}{\bibfnamefont{M.}~\bibnamefont{Rauber}}, \bibnamefont{and}
  \bibinfo{author}{\bibfnamefont{R.}~\bibnamefont{Neumann}},
  \bibinfo{journal}{Nanotechnology} \textbf{\bibinfo{volume}{20}},
  \bibinfo{pages}{325706} (\bibinfo{year}{2009}).

\bibitem[{\citenamefont{Yi{\u{g}}en et~al.}(2013)}]{Yigen:2013we}
\bibinfo{author}{\bibfnamefont{S.} \bibnamefont{Yi{\u{g}}en}},
\bibinfo{author}{\bibfnamefont{V.} \bibnamefont{Tayari}},
\bibinfo{author}{\bibfnamefont{J.~O.} \bibnamefont{Island}},
\bibinfo{author}{\bibfnamefont{J.~M.} \bibnamefont{Porter}}, \bibnamefont{and}
  \bibinfo{author}{\bibfnamefont{A.~R.}~\bibnamefont{Champagne}},
  \bibinfo{journal}{Phys. Rev. B} \textbf{\bibinfo{volume}{87}},
  \bibinfo{pages}{241411} (\bibinfo{year}{2013}).




\bibitem[{\citenamefont{Fong and Schwab}(2012)}]{Fong:2012ut}
\bibinfo{author}{\bibfnamefont{K.~C.} \bibnamefont{Fong}} \bibnamefont{and}
  \bibinfo{author}{\bibfnamefont{K.}~\bibnamefont{Schwab}},
  \bibinfo{journal}{Phys. Rev. X} \textbf{\bibinfo{volume}{2}},
  \bibinfo{pages}{031006} (\bibinfo{year}{2012}).

\bibitem[{\citenamefont{Dicke}(1946)}]{Dicke:1946vp}
\bibinfo{author}{\bibfnamefont{R.~H.} \bibnamefont{Dicke}},
  \bibinfo{journal}{Rev. Sci. Inst.}
  \textbf{\bibinfo{volume}{17}}, \bibinfo{pages}{268} (\bibinfo{year}{1946}).

\bibitem[{\citenamefont{Wellstood et~al.}(1994)\citenamefont{Wellstood, Urbina,
  and Clarke}}]{Wellstood:1994ec}
\bibinfo{author}{\bibfnamefont{F.}~\bibnamefont{Wellstood}},
  \bibinfo{author}{\bibfnamefont{C.}~\bibnamefont{Urbina}}, \bibnamefont{and}
  \bibinfo{author}{\bibfnamefont{J.}~\bibnamefont{Clarke}},
  \bibinfo{journal}{Phys. Rev. B} \textbf{\bibinfo{volume}{49}},
  \bibinfo{pages}{5942} (\bibinfo{year}{1994}).

\bibitem[{\citenamefont{Chen and Clerk}(2012)}]{Chen:2012et}
\bibinfo{author}{\bibfnamefont{W.}~\bibnamefont{Chen}} \bibnamefont{and}
  \bibinfo{author}{\bibfnamefont{A.}~\bibnamefont{Clerk}},
  \bibinfo{journal}{Phys. Rev. B} \textbf{\bibinfo{volume}{86}},
  \bibinfo{pages}{125443} (\bibinfo{year}{2012}).

\bibitem[{Bri()}]{Brian}
\bibinfo{note}{Brian Skinner, unpublished.}



\bibitem[{\citenamefont{Mahan and Bartkowiak}(1999)}]{Mahan:1999eg}
\bibinfo{author}{\bibfnamefont{G.~D.} \bibnamefont{Mahan}} \bibnamefont{and}
  \bibinfo{author}{\bibfnamefont{M.} \bibnamefont{Bartkowiak}},
  \bibinfo{journal}{Appl. Phys. Letts.} \textbf{\bibinfo{volume}{74}},
  \bibinfo{pages}{953} (\bibinfo{year}{1999}).
  
\bibitem[{\citenamefont{Gundrum et~al.}(2005)\citenamefont{Gundrum, Cahill, and
  Averback}}]{Gundrum:2005cx}
\bibinfo{author}{\bibfnamefont{B.~C.} \bibnamefont{Gundrum}},
  \bibinfo{author}{\bibfnamefont{D.~G.} \bibnamefont{Cahill}},
  \bibnamefont{and} \bibinfo{author}{\bibfnamefont{R.~S.}
  \bibnamefont{Averback}}, \bibinfo{journal}{Phys. Rev. B}
  \textbf{\bibinfo{volume}{72}}, \bibinfo{pages}{245426}
  (\bibinfo{year}{2005}).

\bibitem[{\citenamefont{Grosse et~al.}(2011)}]{Grosse:2011ka}
\bibinfo{author}{\bibfnamefont{K.~L.}~\bibnamefont{Grosse}},
  \bibinfo{author}{\bibfnamefont{M.-H.}~\bibnamefont{Bae}},
  \bibinfo{author}{\bibfnamefont{F.} \bibnamefont{Lian}},
  \bibinfo{author}{\bibfnamefont{E.}~\bibnamefont{Pop}}, \bibnamefont{and}
  \bibinfo{author}{\bibfnamefont{W.~P.}~\bibnamefont{King}},
  \bibinfo{journal}{Nat. Nanotechnol.} \textbf{\bibinfo{volume}{6}},
  \bibinfo{pages}{287} (\bibinfo{year}{2011}).

\bibitem[{\citenamefont{Vera-Marun et~al.}(2016)}]{VeraMarun:2016ih}
\bibinfo{author}{\bibfnamefont{I.~J.}~\bibnamefont{Vera-Marun}},
  \bibinfo{author}{\bibfnamefont{J.~J.}~\bibnamefont{van den Berg}},
  \bibinfo{author}{\bibfnamefont{F.~K.}~\bibnamefont{Dejene}}, \bibnamefont{and}
  \bibinfo{author}{\bibfnamefont{B.~J.}~\bibnamefont{van Wees}},
  \bibinfo{journal}{Nat. Commun.} \textbf{\bibinfo{volume}{7}},
  \bibinfo{pages}{11525} (\bibinfo{year}{2016}).
  
  
  

\bibitem[{\citenamefont{Wang et~al.}(2013)\citenamefont{Wang, Meric, Huang,
  Gao, Gao, Tran, Taniguchi, Watanabe, Campos, Muller et~al.}}]{Wang:2013ch}
\bibinfo{author}{\bibfnamefont{L.}~\bibnamefont{Wang}},
  \bibinfo{author}{\bibfnamefont{I.}~\bibnamefont{Meric}},
  \bibinfo{author}{\bibfnamefont{P.~Y.} \bibnamefont{Huang}},
  \bibinfo{author}{\bibfnamefont{Q.}~\bibnamefont{Gao}},
  \bibinfo{author}{\bibfnamefont{Y.}~\bibnamefont{Gao}},
  \bibinfo{author}{\bibfnamefont{H.}~\bibnamefont{Tran}},
  \bibinfo{author}{\bibfnamefont{T.}~\bibnamefont{Taniguchi}},
  \bibinfo{author}{\bibfnamefont{K.}~\bibnamefont{Watanabe}},
  \bibinfo{author}{\bibfnamefont{L.~M.} \bibnamefont{Campos}},
  \bibinfo{author}{\bibfnamefont{D.~A.} \bibnamefont{Muller}},
    \bibinfo{author}{\bibfnamefont{J.} \bibnamefont{Guo}},
      \bibinfo{author}{\bibfnamefont{P.} \bibnamefont{Kim}},
        \bibinfo{author}{\bibfnamefont{J.} \bibnamefont{Hone}},
          \bibinfo{author}{\bibfnamefont{K.~L.} \bibnamefont{Shepard}},
            \bibinfo{author}{\bibfnamefont{C.~R.} \bibnamefont{Dean}},
\bibinfo{journal}{Science}
  \textbf{\bibinfo{volume}{342}}, \bibinfo{pages}{614} (\bibinfo{year}{2013}).

\bibitem[{\citenamefont{L{\'e}onard and Talin}(2011)}]{Leonard:2011dh}
\bibinfo{author}{\bibfnamefont{F.}~\bibnamefont{L{\'e}onard}} \bibnamefont{and}
  \bibinfo{author}{\bibfnamefont{A.~A.} \bibnamefont{Talin}},
  \bibinfo{journal}{Nat. Nanotechnol.} \textbf{\bibinfo{volume}{6}},
  \bibinfo{pages}{773} (\bibinfo{year}{2011}).

\bibitem[{\citenamefont{Matsuda et~al.}(2010)\citenamefont{Matsuda, Deng, and
  Goddard~III}}]{Matsuda:2010fv}
\bibinfo{author}{\bibfnamefont{Y.}~\bibnamefont{Matsuda}},
  \bibinfo{author}{\bibfnamefont{W.-Q.} \bibnamefont{Deng}}, \bibnamefont{and}
  \bibinfo{author}{\bibfnamefont{W.~A.} \bibnamefont{Goddard~III}},
  \bibinfo{journal}{J. Phys. Chem. C}
  \textbf{\bibinfo{volume}{114}}, \bibinfo{pages}{17845}
  (\bibinfo{year}{2010}).

\end{thebibliography}

\end{document}